\documentclass[twocolumn,showpacs,showkeys,superscriptaddress,aps,floatfix]{revtex4}
\usepackage{amsmath,amssymb,booktabs,verbatim}
\usepackage{graphicx}
\usepackage[colorlinks=true,citecolor=blue,filecolor=blue,linkcolor=blue,urlcolor=blue,pdftex]{hyperref}
\newcommand{\1}[1]{\, \mathrm{#1}} 
\newcommand{\n}[1]{\mathrm{#1}}    
\newcommand{\percent}{\%}

\newcommand{\arxiv}[1]{\href{http://arxiv.org/abs/#1}{\texttt{arXiv:#1}}}

\newcommand{\mpi}{\affiliation{Max-Planck-Institut f\"ur Physik, F\"ohringer Ring 6, D-80805 M\"unchen, Germany}}
\newcommand{\tum}{\affiliation{Physik-Department E15, Technische Universit\"at M\"unchen, D-85747 Garching, Germany}}
\newcommand{\oxford}{\affiliation{Department of Physics, University of Oxford, Oxford OX1 3RH, United Kingdom}}
\newcommand{\tubingen}{\affiliation{Eberhard-Karls-Universit\"at T\"ubingen, D-72076 T\"ubingen, Germany}}
\newcommand{\lngs}{\affiliation{INFN, Laboratori Nazionali del Gran Sasso, I-67010 Assergi, Italy}}
\newcommand{\coimbra}{\affiliation{on leave from: Departamento de Fisica, Universidade de Coimbra, P3004 516 Coimbra, Portugal}}
\newcommand{\deceased}{\affiliation{Deceased}}

\begin{document}
\title{Scintillator Non-Proportionality and Gamma Quenching in CaWO$_4$}
\author{R.~F.~Lang}\email{rafael.lang@mpp.mpg.de}\mpi
\author{G.~Angloher}\mpi
\author{M.~Bauer}\tubingen
\author{I.~Bavykina}\mpi
\author{A.~Bento}\mpi \coimbra
\author{A.~Brown}\oxford
\author{C.~Bucci}\lngs
\author{C.~Ciemniak}\tum
\author{C.~Coppi}\tum
\author{G.~Deuter}\tubingen
\author{F.~von~Feilitzsch}\tum
\author{D.~Hauff}\mpi
\author{S.~Henry}\oxford
\author{P.~Huff}\mpi
\author{J.~Imber}\oxford
\author{S.~Ingleby}\oxford
\author{C.~Isaila}\tum
\author{J.~Jochum}\tubingen
\author{M.~Kiefer}\mpi
\author{M.~Kimmerle}\tubingen
\author{H.~Kraus}\oxford
\author{J.-C.~Lanfranchi}\tum
\author{M.~Malek}\oxford
\author{R.~McGowan}\oxford
\author{V.~B.~Mikhailik}\oxford
\author{E.~Pantic}\mpi
\author{F.~Petricca}\mpi
\author{S.~Pfister}\tum
\author{W.~Potzel}\tum
\author{F.~Pr\"obst}\mpi
\author{S.~Roth}\tum
\author{K.~Rottler}\tubingen
\author{C.~Sailer}\tubingen
\author{K.~Sch\"affner}\mpi
\author{J.~Schmaler}\mpi
\author{S.~Scholl}\tubingen
\author{W.~Seidel}\mpi
\author{L.~Stodolsky}\mpi
\author{A.~J.~B.~Tolhurst}\oxford
\author{I.~Usherov}\tubingen
\author{W.~Westphal}\tum\deceased

\begin{abstract}
We measure and explain scintillator non-proportionality and gamma quenching of $\n{CaWO_4}$ at low energies and low temperatures. Phonons that are created following an interaction in the scintillating crystal at temperatures of $\sim15\1{mK}$ are used for a calorimetric measurement of the deposited energy, and the scintillation light is measured with a separate cryogenic light detector. Making use of radioactivity intrinsic to the scintillating crystal, the scintillator non-proportionality is mapped out to electron energies $<5\1{keV}$. The observed behavior is in agreement with a simple model based on Birks' law and the stopping power $\n{d}E/\n{d}x$ for electrons. We find for Birks' constant $k_B=(18.5\pm0.7)\1{nm/keV}$ in $\n{CaWO_4}$. Gamma lines allow a measurement of the reduced light yield of photons with respect to electrons, as expected in the presence of scintillator non-proportionality. In particular, we show that gamma-induced events in $\n{CaWO_4}$ give only about $90\percent$ of the light yield of electrons, at energies between $40\1{keV}$ and $80\1{keV}$. 
\end{abstract}

\maketitle

\section{Introduction}
Scintillation is among the oldest and most classic of particle detector technologies. For photons and electrons with energies below $\sim 100\1{keV}$, it appears that the light output from the scintillator is not simply proportional to the absorbed energy~\cite{moses2008}. Recently, this so-called scintillator non-proportionality is receiving renewed attention thanks to the utilization of scintillators at these low energies in the search for dark matter~\cite{freedman2003}. For ions and other particles producing heavily ionizing tracks on the one hand, this effect can be understood as a saturation effect that occurs for the light produced per unit length $\n{d}L/\n{d}x$ when an increase in local energy deposition $\n{d}E/\n{d}x$ does not lead to more light. This phenomenon is encoded in Birks' law~\cite{birks1967}
\begin{eqnarray}
  \frac{\n{d}L}{\n{d}x} = \frac{A \: \n{d}E/\n{d}x}{1+ k_B\:\n{d}E/\n{d}x}.\label{eq:birks}
\end{eqnarray}
$A$ and $k_B$ are empirical constants that depend on the material and have to be determined from the data. At high energy loss $\n{d}E/\n{d}x$, the light output is no longer proportional to the energy loss, due to the term $k_B\:\n{d}E/\n{d}x$ becoming larger than unity. Eventually, $\n{d}L/\n{d}x$ becomes independent of $\n{d}E/\n{d}x$ and the total light from an event depends only on the track length~\cite{bavykina2007}.

\begin{figure}[htbp]
\begin{center}\includegraphics[angle=90,width=1\columnwidth,clip,trim=0 0 155 0]{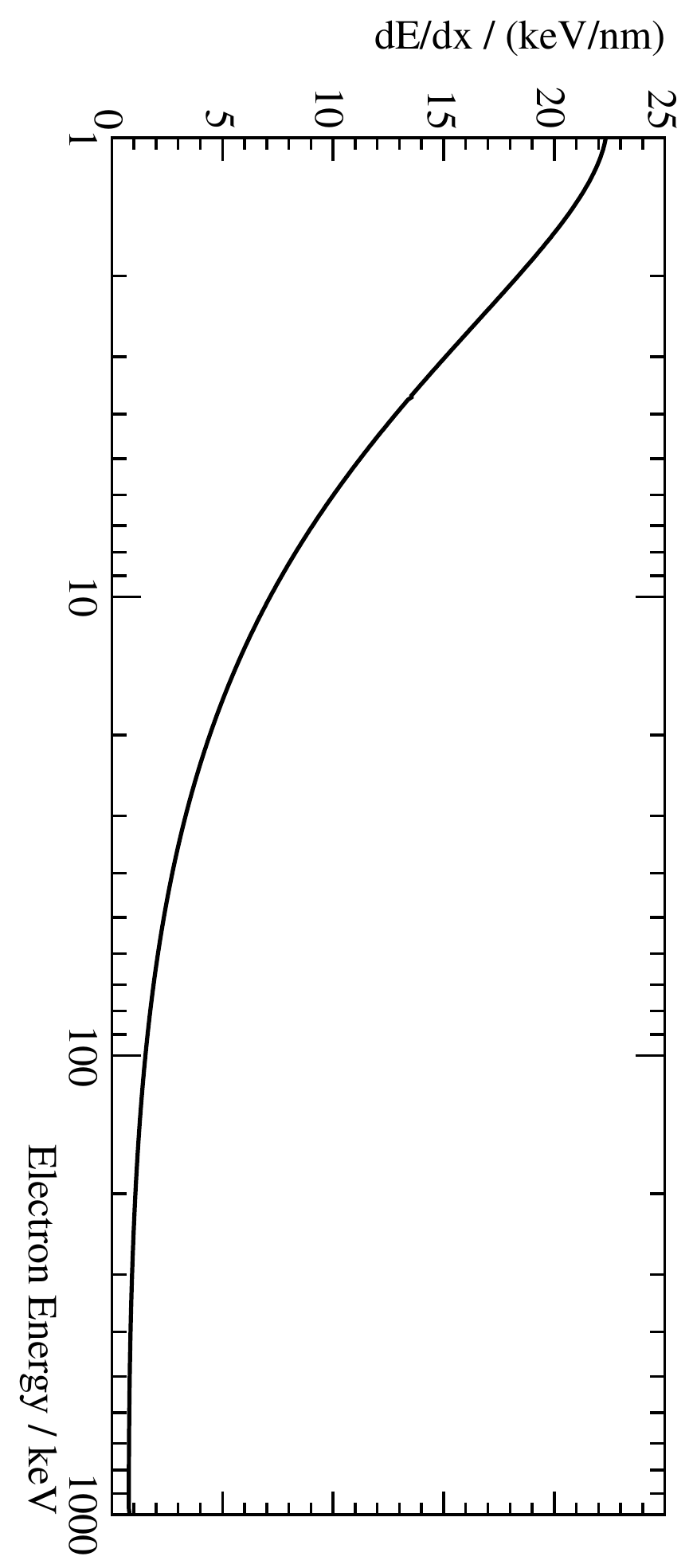}\end{center}
\vspace{-5mm}\caption{Energy dependence of $\n{d}E/\n{d}x$ for electrons in $\n{CaWO_4}$, from the \texttt{estar} database~\cite{berger0001}.}
\label{fig:dedx}\end{figure}

Electrons on the other hand are lightly ionizing, but for low-energy electrons, there is in fact an increase in local energy loss. As an example, figure~\ref{fig:dedx} shows $\n{d}E/\n{d}x\,(E)$ for electrons in $\n{CaWO_4}$, the material we shall study here. One might hence anticipate some effect from the non-proportionality in equation~\ref{eq:birks} also for low-energy electrons, where $\n{d}E/\n{d}x$ for energies near the end of the track (i.e. below $\sim10\1{keV}$) is significantly increased.

Evidently, the most direct way to measure the response of a scintillator to electrons is to study its light output from single monoenergetic electrons, over a range of low energies. Since there are no natural electron sources of this type, the Compton-coincidence technique has been developed~\cite{valentine1994}. In this technique, one uses a monoenergetic gamma with relatively high energy, which undergoes a Compton scattering in the material under study. The scattered photon is observed in a separate detector. Knowledge of the scattered photon's angle leads, via Compton scattering kinematics, to a recoil electron of known energy in the material under study. This method has been employed successfully to map out the response of various scintillators to low-energy electrons (see~\cite{rooney1997} and references therein).

Another method to measure the electron response of a scintillator uses the monochromatic X-ray radiation from a synchrotron source. The scintillation light yield of an excitation above the characteristic K-edge of the element is measured. When the energy independent response from the cascade event that is due to the relaxation of holes in the K-shell is subtracted, it is possible to derive the response due to the K-shell photoelectron alone~\cite{wayne1998,khodyuk2009}.

In this note, we would like to draw attention to the possibility of a new method, which has arisen through the investigation of cryogenic scintillating detectors in the search for dark matter. A two-channel readout allows both the energy and the light yield of an event to be determined. A large sample of low energy electron events with both known energy and light yield, acquired as a background in the search for dark matter, allows a precise measurement of the scintillator response and its non-linearity.

The fundamental processes of excitation of electrons in deep shells of atoms are temperature independent. Indeed, it has already been shown that the non-proportionality of self-activated scintillators such as $\n{CaWO_4}$ or BGO exhibit no temperature dependence~\cite{moszynski2004,moszynski2005} even thought their light yield changes significantly with temperature~\cite{gironnet2008,mikhailik2007}. Hence, measuring at millikelvin temperatures allows one to infer the scintillator non-proportionality also at room temperature.

\section{Setup}
The CRESST-II experiment~\cite{angloher2009} is built to search for and detect a new particle species, so-called weakly interacting massive particles (WIMPs), which could constitute the dark matter known to exist in the universe~\cite{amsler2008b}. The expected interaction rates are less than one event per kilogram of target material, per year of observation and per keV in the energy range between $\sim10\1{keV}$ and $\sim40\1{keV}$~\cite{angloher2009}. This requires a substantial reduction of backgrounds from ambient radioactivity. To this end, the CRESST-II experiment is located in the Laboratori Nazionali del Gran Sasso under an average rock overburden of $1400\1{m}$ to shield the experiment from cosmic rays, and additionally, the cryostat with the target materials is surrounded by thicknesses of $45\1{cm}$ of polyethylene, $20\1{cm}$ of lead, and $14\1{cm}$ of copper.

The remaining activity is mostly due to beta emitters intrinsic to the employed target materials~\cite{lang2009b}. To distinguish the expected WIMP-induced nuclear-recoil signal from this remaining background, an active discrimination technique is used. In CRESST-II, we use scintillating tungstates ($\n{CaWO_4}$, $\n{ZnWO_4}$) as target materials. These crystals are shaped as cylinders with a diameter of $4\1{cm}$ and similar height, thus weighing about $300\1{g}$ ($\n{CaWO_4}$) or $400\1{g}$ ($\n{ZnWO_4}$) each. They are cooled to temperatures of $\sim 15\1{mK}$ and operated as cryogenic calorimeters~\cite{angloher2005}: non-thermal phonons generated in a particle interaction are collected by a superconducting phase-transition thermometer~\cite{meier1999} which is stabilized in its transition to the superconducting state by means of an additional heater structure on the thermometer~\cite{angloher2005}. By this configuration, the resulting change of resistance following a particle interaction can be read out with a SQUID-based readout scheme~\cite{seidel1990,henry2007} and is used for a calorimetric measurement of the interaction energy.

An additional cryogenic detector~\cite{petricca2004} in the vicinity of the crystal is used to detect the scintillation light, typically about $1\percent$ of the total energy for electrons~\cite{westphal2006}. The light detector response is calibrated and linearized independently of the scintillation light using externally created pulses injected into a heater structure on the thermometer~\cite{angloher2005}. Hence, each particle interaction results in two parameters: the energy $E$ of the interaction as measured directly with the calorimeter, and the light yield $\frac{L}{E}$, which is defined as the ratio of detected light $L$ over the energy $E$ of the interaction. We measure $\frac{L}{E}$ in units of $\n{keV_{ee}/keV}$, where $\n{keV_{ee}}$ (keV electron-equivalent) is defined via a $\n{{}^{57}Co}$ calibration source such that $E=122\1{keV}$ gamma rays produce photoelectrons with a light yield of $\frac{L}{E}=1\1{keV_{ee}/keV}$.

Relevant in the search for dark matter is the possibility to discriminate electron or gamma-induced events from nuclear recoils based on a strongly reduced light yield of the latter. For example, recoiling alphas have $18\percent$ of the electron light yield. Due to coherence effects, dark matter induced events are expected to mainly cause recoiling tungsten nuclei, which show less than $2.5\percent$ of the electron light yield~\cite{bavykina2007,ninkovic2006}. The effects discussed in this note are more subtle and concern slight variations within the population of electron and gamma-induced events.

\section{Scintillator Non-Proportionality}

\begin{figure}[htbp]
\begin{center}\includegraphics[angle=90,width=1\columnwidth]{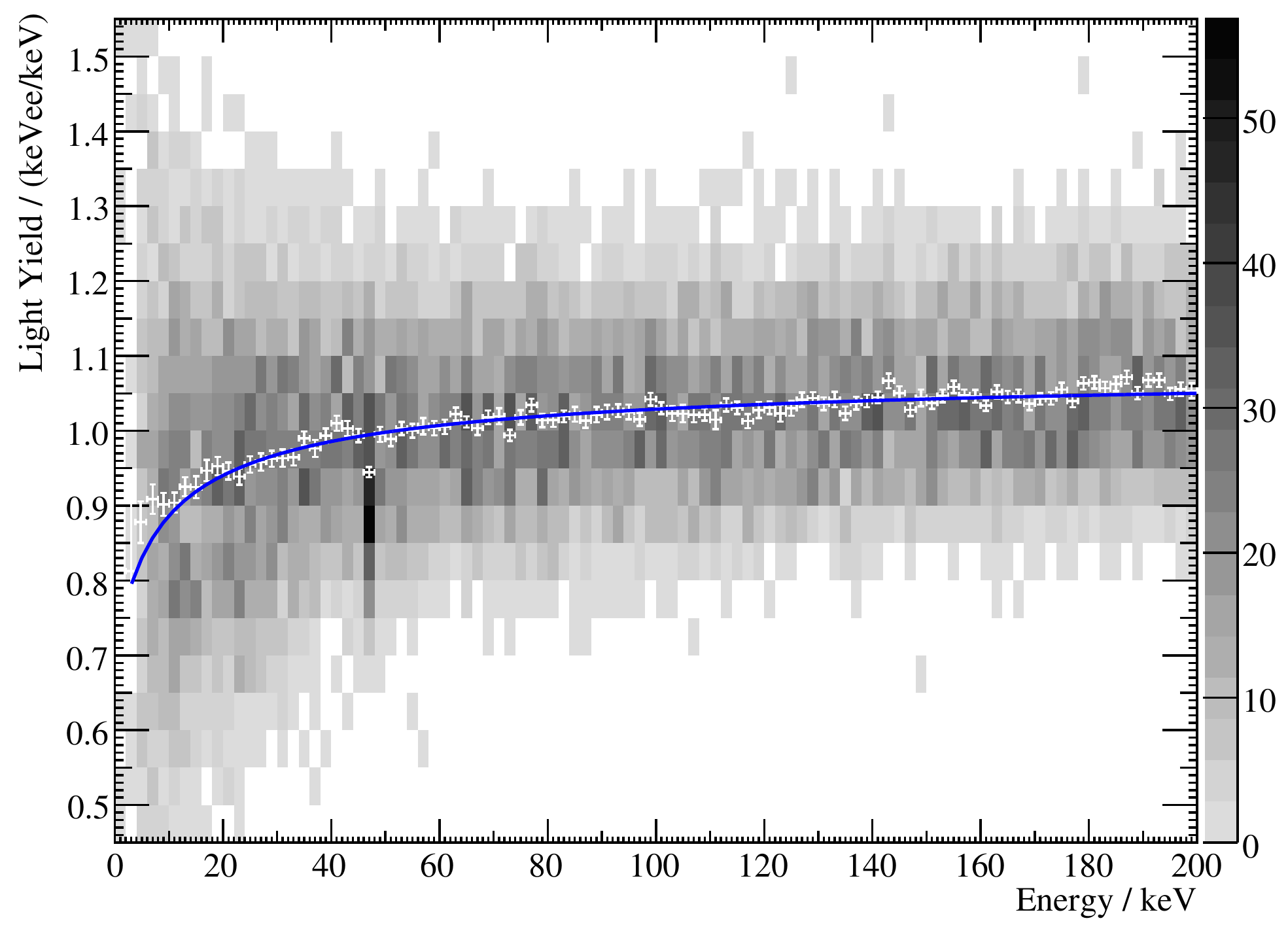}\end{center}
\vspace{-5mm}\caption{Electron and gamma events observed with one crystal after an exposure of $12.31\1{kg\,d}$ (detector \textsc{Daisy}/run 27). Number of entries per bin are according to the grey scale on the right. The white data points mark the mean of the light yield in each $2\1{keV}$ energy bin together with the error bars of the mean and the width of the energy bin; the scintillator non-proportionality is clearly visible. The solid (blue) line is a fit according to the model described in the text.}
\label{fig:DaisyRun27E}\end{figure}

The data presented here was taken over a period of about one month while no calibration source was present. The accumulated data is shown in the light yield-energy plane $\frac{L}{E}\,(E)$ in figure~\ref{fig:DaisyRun27E} for interaction energies $E$ up to $200\1{keV}$. The events are mainly from a continuous electron background due to an internal contamination of the crystal with $\n{{}^{90}Sr}$ and other beta emitters~\cite{lang2009b}. In addition, a few gamma lines of both internal and external origins are superimposed; we will come back to these in section~\ref{sec:gammaquenching}.

Also shown in figure~\ref{fig:DaisyRun27E} is the mean of the light yield in each $2\1{keV}$ energy bin, which clearly decreases with decreasing energy. This scintillator non-proportionality has previously been measured for $\n{CaWO_4}$ with small crystals under external gamma irradiation~\cite{moszynski2005} at room and liquid nitrogen temperatures. Measurements with other tungstates at room temperature also show the same effect~\cite{dorenbos1995,kraus2009}.

This scintillator non-proportionality can be explained simply by the stopping power per unit path length $\n{d}E/\n{d}x\,(E)$ for electrons in $\n{CaWO_4}$, together with Birks' law (equation~\ref{eq:birks}). Integrating Birks' law allows a calculation of the light yield $\frac{L}{E}\,(E)$:
\begin{eqnarray}
	\frac{L}{E}\,(E) = \frac{1}{E} \int_0^E \frac{\n{d}L}{\n{d}E'} \,\n{d}E'
                    = \frac{1}{E} \int_0^E \frac{\n{d}L}{\n{d}x} \left( \frac{\n{d}E'}{\n{d}x} \right)^{-1} \,\n{d}E'.\label{eq:model}
\end{eqnarray}

For $\n{d}E/\n{d}x\,(E)$, we use the data from the \texttt{estar} database~\cite{berger0001} shown in figure~\ref{fig:dedx}. With $\n{d}L/\n{d}x$ from equation~\ref{eq:birks}, we can perform a fit of equation~\ref{eq:model} to our data, which yields the parameters $A$ and $k_B$. This fit is also shown in figure~\ref{fig:DaisyRun27E}. When the errors stated by the \texttt{estar} database are taken into account, the variation of the fit is within the thickness of the line in figure~\ref{fig:DaisyRun27E}, even when allowing for a $\pm20\percent$ variation for energies $<10\1{keV}$. The model can be seen to describe the observed scintillator non-proportionality well.

From the fit we obtain the Birks constants as $A=(1.096\pm0.003)\1{keV_{ee}/keV}$ and $k_B=(18.5\pm0.7)\1{nm/keV}$ for electrons in our $\n{CaWO_4}$ crystal. Since $A$ describes a linear dependence between $\n{d}L/\n{d}x$ and $\n{d}E/\n{d}x$, as appropriate for higher energies, the value $A\approx1\1{keV_{ee}/keV}$ was to be expected by construction of the $\n{keV_{ee}}$ unit. The parameter $k_B$ characterizes at what $\n{d}E/\n{d}x$ the light yield $\frac{L}{E}$ begins to saturate, and $1/k_B$ may be thought of as a quantity $\left(\n{d}E/\n{d}x\right)_{\n{saturation}}$. Dividing by the density of the $\n{CaWO_4}$ crystal of $6.134\1{g/cm^3}$, we have $\left(\n{d}E/\n{d}x\right)_{\n{saturation}}=(88\pm3)\1{MeV\,cm^2/g}$. This is consistent with the range $8\1{MeV\,cm^2/g} < \left(\n{d}E/\n{d}x\right)_{\n{saturation}} < 160\1{MeV\,cm^2/g}$ which was estimated in~\cite{bavykina2007} based on a different method.

\section{Gamma Quenching}\label{sec:gammaquenching}

The absorption of a gamma is typically a complicated process, where an electron is ejected from its shell, leaving a vacancy. The electron gives rise to scintillation, but has less energy than the incoming gamma. The binding energy of the electron will be released in the form of an Auger electron (mainly for light elements) or X-rays (for heavy elements), which can in turn eject other electrons, leading to an electron cascade for each gamma traversing the scintillator. In this way, the final energy transfer to the material following a gamma interaction takes place through a number of low-energy electrons, further enhancing the role of the last few keV of electron tracks~\cite{rooney1997} (compare figure~\ref{fig:dedx}).

Hence, in the presence of scintillator non-proportionality, we can expect a reduced light yield for low-energy gamma events with respect to electron events of the same energy, since the initial energy of the gamma is distributed over many electrons. Given a parametrization of the scintillator non-proportionality for electrons, the light yield for gammas has been calculated in the framework of this model for a few materials~\cite{rooney1997}, although a calculation for $\n{CaWO_4}$ is not known to us. It thus appears possible that a non-linear gamma response of the light yield can be understood via the saturation effect of equation~\ref{eq:birks}.

Here, we demonstrate the presence of a reduced light yield for gamma events in $\n{CaWO_4}$ crystals. It is already visible in figure~\ref{fig:DaisyRun27E} that the $46.5\1{keV}$ line has a somewhat reduced light yield with respect to the continuous electron background. In addition, at $122\1{keV}$ where the $\n{keV_{ee}}$ unit is normalized to unity with gamma events from a $\n{{}^{57}Co}$ calibration, the mean of the light yield of the observed electron background is $(1.03\pm0.01)\1{keV_{ee}}$, pointing to a reduced light yield of the gamma events in the calibration. 

\begin{figure}[htbp]
\begin{center}\includegraphics[angle=90,width=1\columnwidth,clip,trim=0 0 140 0]{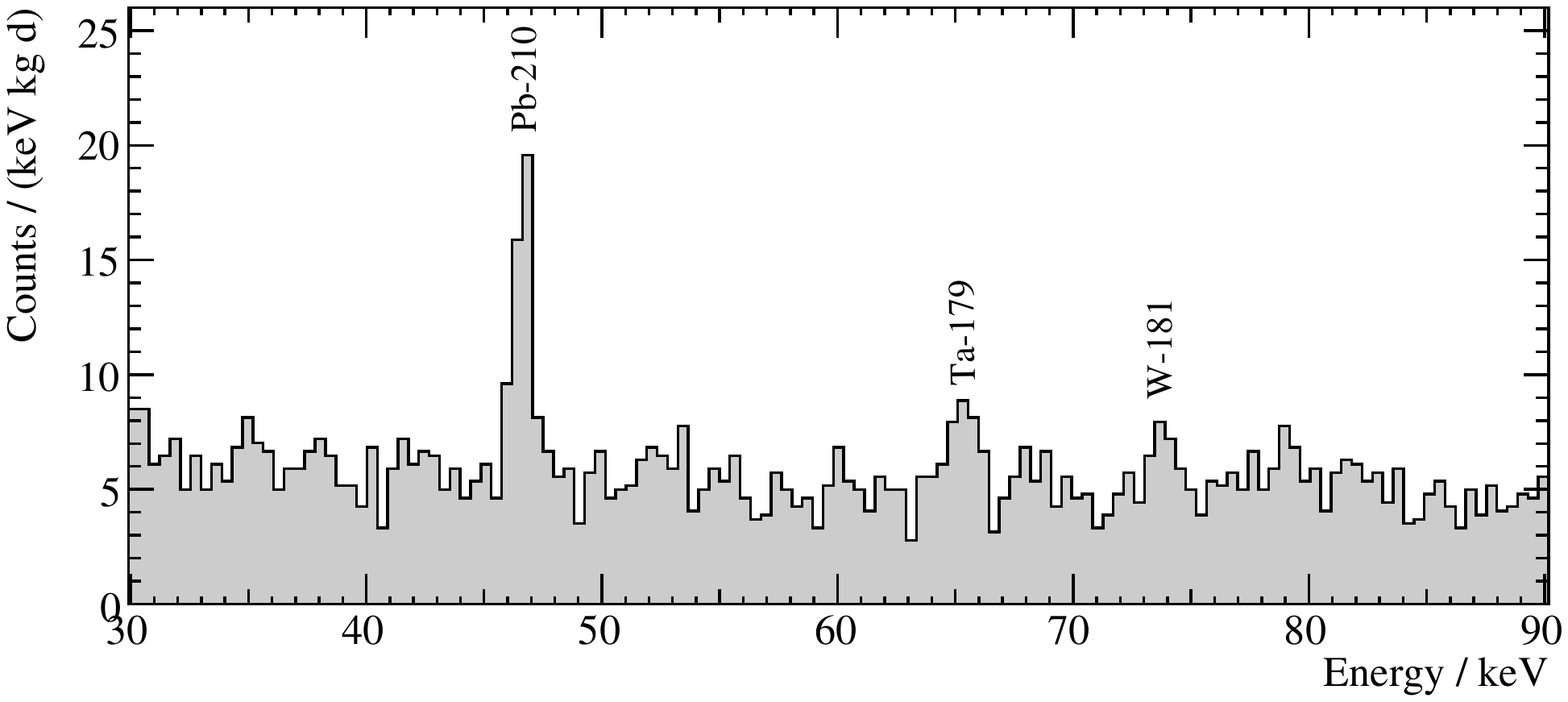}\end{center}
\vspace{-5mm}\caption{Spectrum of events observed in figure~\ref{fig:DaisyRun27E}. The flat background is due to the beta spectrum of $\n{{}^{90}Sr}$. The three observed lines are labeled according to their origin: one is due to an external contamination with $\n{{}^{210}Pb}$, the other two are due to cosmogenic activation of the tungsten.}
\label{fig:SpectrumDaisyRun27}\end{figure}

A quantitative analysis needs to disentangle the light yield of gamma events from that of the electron background. To this end, figure~\ref{fig:SpectrumDaisyRun27} shows the spectrum of the data shown in figure~\ref{fig:DaisyRun27E}, in which three gamma lines can be identified. One at $46.5\1{keV}$ is due to an external contamination with $\n{{}^{210}Pb}$. Two more lines at $65.4\1{keV}$ and $73.7\1{keV}$ are due to tungsten activated by cosmic radiation. The remaining spectrum is rather flat and dominated by the beta spectrum of $\n{{}^{90}Sr}$~\cite{lang2009b}.

\begin{figure}[htbp]
\begin{center}\includegraphics[angle=90,width=1\columnwidth]{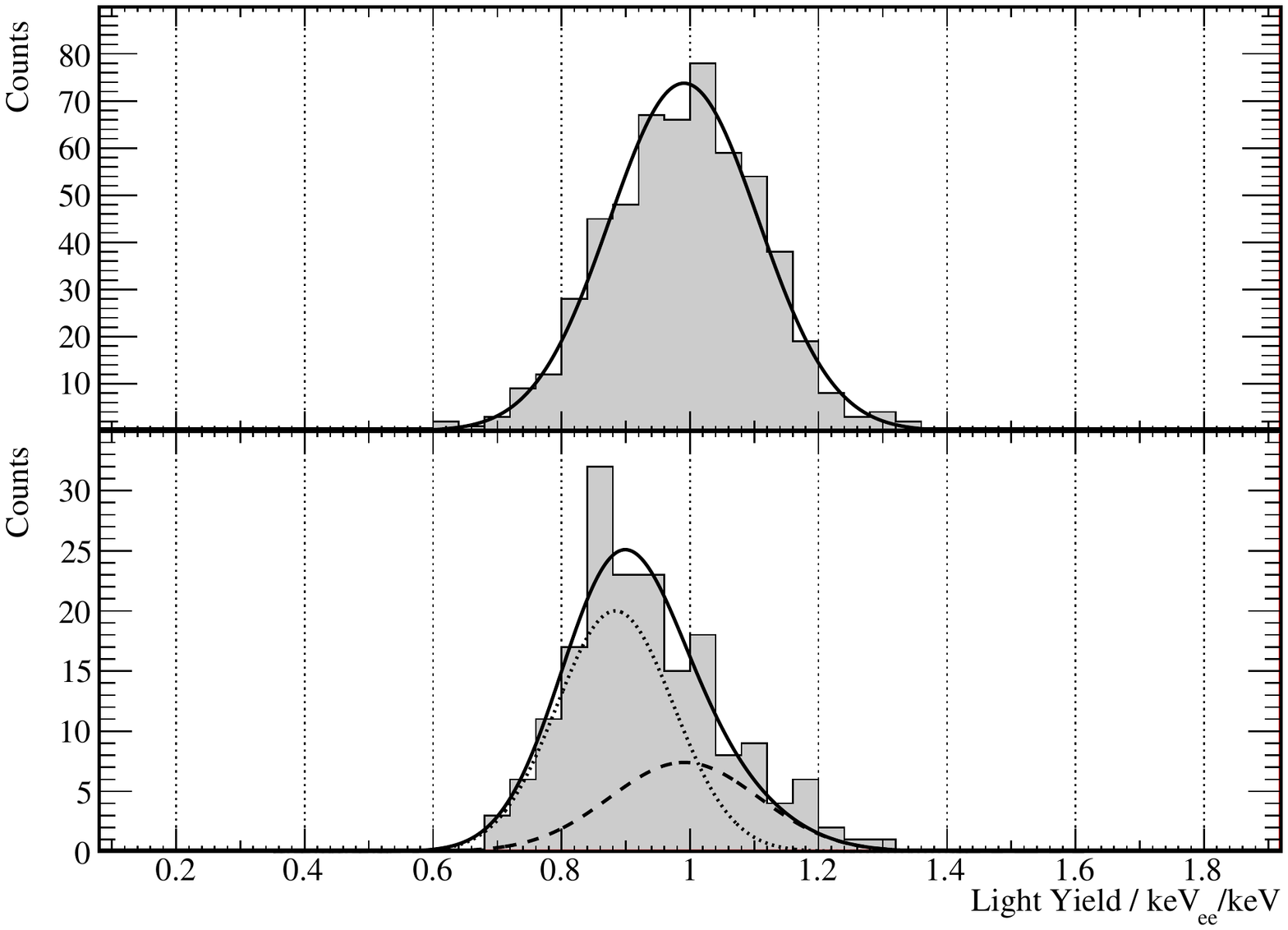}\end{center}
\vspace{-5mm}\caption{Upper histogram: fit of a Gaussian to the off-peak events (only electron events) around the $\n{{}^{210}Pb}$ peak. Resulting fit parameters are 
$y_{\n{off}}=(0.991\pm0.005)\1{keV_{ee}/keV}$ for the light yield (in agreement with our model which gives $0.994\1{keV_{ee}/keV}$), and $\sigma_{\n{off}}=(0.116\pm0.004)\1{keV_{ee}/keV}$. Lower histogram: fit of the sum of two Gaussians (full line) to the $\n{{}^{210}Pb}$ on-peak distribution. One Gaussian (dashed) has all parameters taken from the off-peak distribution and represents the electron events. All parameters of the second Gaussian (dotted) which accounts for the gamma events are left free, and the fit finds 
$y_{\n{on}}=(0.89\pm0.02)\1{keV_{ee}/keV}$ and $\sigma_{\n{on}}=(0.09\pm0.01)\1{keV_{ee}/keV}$.}
\label{fig:QuenchingPb2e}\end{figure}

We proceed as follows: for each of the observed gamma lines at energy $E_{\gamma}$, a Gaussian is fitted to the light yield distribution of \textit{off-peak} events, which are taken from the intervals $[E_{\gamma}-5\1{keV},E_{\gamma}-1\1{keV}]$ and $[E_{\gamma}+1\1{keV},E_{\gamma}+5\1{keV}]$. A second population is taken \textit{on-peak}, from $[E_{\gamma}-0.4\1{keV},E_{\gamma}+0.4\1{keV}]$, which corresponds to a $2\sigma$ interval around the line, given the energy resolution of the calorimeter at these energies. The sum of two Gaussians is fitted to the on-peak distribution: one Gaussian is completely fixed to have the same mean light yield $y$ and the same width $\sigma$ as the off-peak Gaussian, and its amplitude $A$ is scaled to the smaller on-peak energy bin width. For the second Gaussian all parameters are left free. Figure~\ref{fig:QuenchingPb2e} shows the off- and on-peak distributions and the fits for the $\n{{}^{210}Pb}$ line at $46.5\1{keV}$. Clearly, the additional gamma events from the $\n{{}^{210}Pb}$ line show a reduced light yield compared to electrons of the same energy of only $(0.89\pm0.02)\1{keV_{ee}/keV}/(0.991\pm0.005)\1{keV_{ee}/keV}=(0.90\pm0.02)$ of the electron yield.

Since $46.5\1{keV}$ gammas are absorbed within the first millimeter in $\n{CaWO_4}$, an alternative explanation of the reduced gamma light yield could be based on the following argument: the crystals are cut in cylindrical shape to have the largest crystals possible in the search for dark matter. Hence, events happening close to the mantle surface could show a reduced light yield due to geometrical light trapping in the crystal, following total internal reflections on the mantle surface, in which case the $\n{{}^{210}Pb}$ line could show a reduced light yield simply due to light trapping. However, this is not the case, as we demonstrate considering sources of photons which are uniformly distributed throughout the crystal.

Exposition to cosmic radiation of the crystals or the raw materials from which they are grown leads to an activation of the tungsten. We observe products from the reactions $\n{{}^{182}W(p,\alpha){}^{179}Ta}$ and $\n{{}^{183}W(p,t){}^{181}W}$. The resulting isotopes will be distributed homogeneously throughout the whole crystal, independently of whether the activation takes place before or after crystal growth.

\begin{figure}[htbp]
\begin{center}\includegraphics[angle=90,width=1\columnwidth]{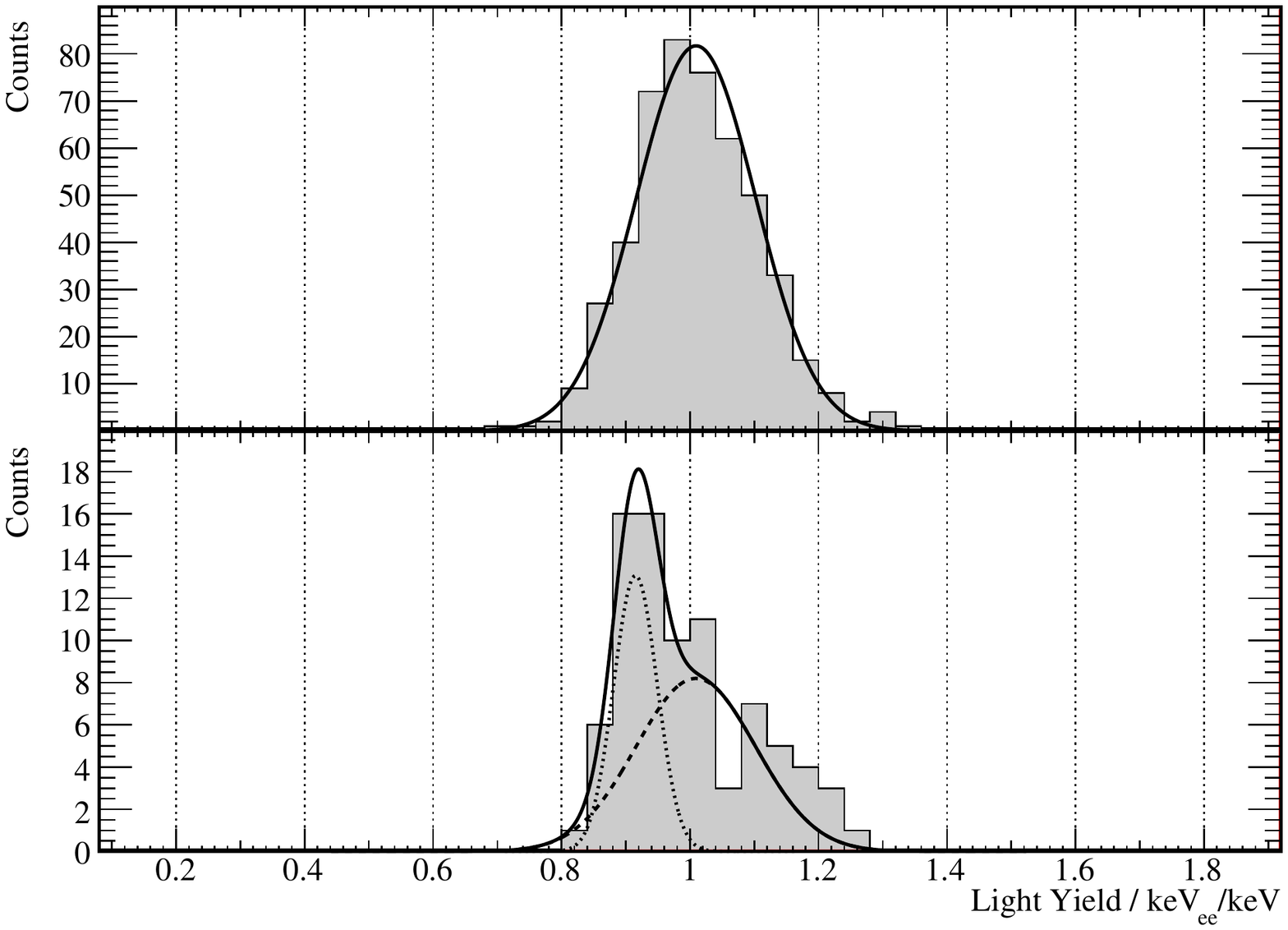}\end{center}
\vspace{-5mm}\caption{The Gaussian fit to the off-peak events around the $\n{{}^{179}Ta}$ peak (upper histogram, only electron events) gives 
$y_{\n{off}}=(1.009\pm0.004)\1{keV_{ee}/keV}$ for the light yield (in agreement with our model which gives $1.011\1{keV_{ee}/keV}$) and $\sigma_{\n{off}}=(0.093\pm0.003)\1{keV_{ee}/keV}$. Fitting the sum of two Gaussians to the on-peak distribution (lower histogram), one Gaussian representing electron events (dashed), the other gamma events (dotted), gives  
$y_{\n{on}}=(0.92\pm0.01)\1{keV_{ee}/keV}$, and $\sigma_{\n{on}}=(0.034\pm0.008)\1{keV_{ee}/keV}$ for the light yield of gamma events.}
\label{fig:QuenchingTa2e}\end{figure}

The isotope $\n{{}^{179}Ta}$ decays with a half-life of $1.8\1{years}$ in an electron capture process into $\n{{}^{179}Hf}$~\cite{firestone1996}. In most cases, this results in a vacancy in the K-shell,
which is quickly filled, eventually leading to a cascade of low-energy X-rays. In our calorimetric measurement we observe the full binding energy of a hafnium K-shell electron, namely $65.4\1{keV}$. The corresponding off- and on-peak distributions together with the Gaussian fits are shown in figure~\ref{fig:QuenchingTa2e}. The resulting gamma light yield is $(0.92\pm0.01)\1{keV_{ee}/keV}/(1.009\pm0.004)\1{keV_{ee}/keV}=(0.91\pm0.01)$ of the electron light yield, just as in the case of $\n{{}^{210}Pb}$ events. Thus, light trapping can be excluded as explanation for the reduced gamma light yield.

A consistent observation can be made with a second line from cosmic activation: the isotope $\n{{}^{181}W}$ decays with a half-life of $121\1{days}$ in an electron capture process into $\n{{}^{181}Ta}$ under emission of an additional $6.2\1{keV}$ gamma~\cite{firestone1996}. We observe the line at an energy corresponding to the binding energy of a tantalum K-shell electron ($67.4\1{keV}$) plus the energy of the gamma, namely $73.7\1{keV}$. The light yield of this line is derived as above, and again the light yield of gammas is found to be reduced to $0.88\pm0.01$ of the light yield for electron events of the same energy. 

\section{Conclusion}

We have presented measurements of scintillator non-proportionality for low-energy electrons in $\n{CaWO_4}$. The crystal under investigation is cooled to mK temperatures, and the non-thermal phonon signal is measured with a superconducting phase-transition thermometer for a calorimetric measurement of the interaction energy. The light yield is measured with a separate cryogenic light detector. The observed scintillator non-proportionality for low-energy electrons is well described by a simple model based on Birks' law and the stopping power $\n{d}E/\n{d}x$ of electrons in $\n{CaWO_4}$. Gamma events show a light yield that is reduced to $\sim90\percent$ of the light yield of electron events with the same energy. This is understood within the same model, where gamma events in a scintillator cause a cascade of low-energy electrons, which in turn produce the scintillation light. Our findings help to identify nuclear recoil signals in experiments that use $\n{CaWO_4}$ to search for dark matter.

\section{Acknowledgments}

This work was partially supported by funds of the DFG (SFB 375 and Transregio 27 ``Neutrinos and Beyond''), the Munich Cluster of Excellence (``Origin and Structure of the Universe''), the EU networks for Cryogenic Detectors (ERB-FMRXCT980167) and for Applied Cryogenic Detectors (HPRN-CT2002-00322), and the Maier-Leibnitz-Laboratorium (Garching). Support was provided by the Science and Technology Facilities Council.


\begin{thebibliography}{00}
\bibitem{moses2008}
W.~W.~Moses et~al., IEEE Transactions on Nuclear Science \textbf{55}, 1049 (2008).
\bibitem{freedman2003}
W.~L.~Freedman and M.~S.~Turner, Review of Modern Physics \textbf{75}, 1433 (2003), \arxiv{astro-ph/0308418}.
\bibitem{birks1967}
J.~B.~Birks, \textit{Theory and Practice of Scintillation Counting}, Pergamon, 1967.
\bibitem{bavykina2007}
I.~Bavykina et~al., Astroparticle Physics \textbf{28}, 489 (2007), \arxiv{0707.0766}.
\bibitem{berger0001}
M.~J.~Berger et~al., \textit{Stopping-Power and Range Tables for Electrons, Protons, and Helium Ions}, version 1.2.3, 
  \url{http://www.physics.nist.gov/PhysRefData/Star/Text/ESTAR.html}, 2005.
\bibitem{valentine1994}
J.~D.~Valentine and B.~D.~Rooney, Nuclear Instruments and Methods A \textbf{353}, 37 (1994).
\bibitem{rooney1997}
B.~D.~Rooney and J.~D.~Valentine, IEEE Transactions on Nuclear Science \textbf{44}, 509 (1997).
\bibitem{wayne1998}
L.~R.~Wayne et~al., Nuclear Instruments and Methods A \textbf{411} 351 (1998).
\bibitem{khodyuk2009}
I.~V.~Khodyuk, J.~T.~M.~de~Haas and P.~Dorenbos, submitted to IEEE Transactions on Nuclear Science.
\bibitem{moszynski2005}
M.~Moszy\'nski et~al., Nuclear Instruments and Methods A \textbf{553}, 578 (2005).
\bibitem{moszynski2004}
M.~Moszy\'nski et~al., IEEE Transactions on Nuclear Science \textbf{51} 1074 (2004).
\bibitem{mikhailik2007}
V.~B.~Mikhailik, H.~Kraus, S.~Henry and A.~J.~B.~Tolhurst, Physical Review B \textbf{75} 184308 (2007).
\bibitem{gironnet2008}
J.~Gironnet et~al., Nuclear Instruments and Methods A \textbf{594} 358 (2008).
\bibitem{angloher2009}
G.~Angloher et~al. (The CRESST Collaboration), Astroparticle Physics \textbf{31}, 270 (2009), \arxiv{0809.1829}.
\bibitem{amsler2008b}
C.~Amsler et~al., Physics Letters B \textbf{667} 1 (2008), available from \url{http://pdg.lbl.gov/}, see in particular the \textit{Dark Matter} review.
\bibitem{lang2009b}
R.~F.~Lang et~al., accepted for publication in Astroparticle Physics, \arxiv{0905.4282}.
\bibitem{angloher2005}
G.~Angloher et~al. (The CRESST Collaboration), Astroparticle Physics \textbf{23}, 325 (2005), \arxiv{astro-ph/0408006}.
\bibitem{meier1999}
O.~Meier et~al., Superconductor Science and Technology \textbf{12}, 1033 (1999).
\bibitem{seidel1990}
W.~Seidel et~al., Physics Letters B \textbf{236}, 483 (1990).
\bibitem{henry2007}
S.~Henry et~al., Journal of Instrumentation \textbf{2}, 11003 (2007).
\bibitem{petricca2004}
F.~Petricca et~al., Nuclear Instruments and Methods A \textbf{520}, 193 (2004).
\bibitem{westphal2006}
W.~Westphal et~al., Nuclear Instruments and Methods A \textbf{559}, 372 (2006).
\bibitem{ninkovic2006}
J.~Ninkovi\'c et~al., Nuclear Instruments and Methods A \textbf{564}, 567 (2006), \arxiv{astro-ph/0604094}.
\bibitem{dorenbos1995}
P.~Dorenbos, J.~T.~M.~de~Haas and C.~W.~E.~van~Eijk, IEEE Transactions on Nuclear Science \textbf{42}, 2190 (1995).
\bibitem{kraus2009}
H.~Kraus et~al., Nuclear Instruments and Methods A \textbf{600}, 594 (2009).
\bibitem{firestone1996}
R.~B.~Firestone et~al., \textit{Table of Isotopes (CD ROM Edition)}, 1st ed., John Wiley \& Sons, New York, 1996.
\end{thebibliography}
\end{document}